# Designing Informative Securities[*]

**Yiling Chen**
Harvard University

**Mike Ruberry**
Harvard University

**Jennifer Wortman Vaughan**
University of California, Los Angeles

## Abstract

We create a formal framework for the design of informative securities in prediction markets. These securities allow a market organizer to infer the likelihood of events of interest as well as if he knew all of the traders' private signals. We consider the design of markets that are always informative, markets that are informative for a particular signal structure of the participants, and informative markets constructed from a restricted selection of securities. We find that to achieve informativeness, it can be necessary to allow participants to express information that may not be directly of interest to the market organizer, and that understanding the participants' signal structure is important for designing informative prediction markets.

## 1 INTRODUCTION

Prediction markets are often used to better understand the likelihood of future events. Consider, for example, a market that predicts whether a particular candidate will become the U.S. President. Traders in the market may have diverse private information, like whether the candidate will win a particular state or receive millions of dollars in donations. Pooling this information could lead to an accurate forecast of the likelihood that the candidate will be elected, but may or may not be possible depending on how well the market is designed.

A typical prediction market offers securities with payoffs associated with future events. For example, a prediction market might offer a security worth $1 if the candidate is elected and $0 otherwise. A risk neutral trader who believes that the probability of election is $p$ would be willing to buy this security at any price less than $\$p$, and (short) sell the security at any price above $\$p$. For this reason, the market price of a security of this form is frequently interpreted as the traders' collective belief about the likelihood of election.

Prediction markets have been shown to produce forecasts at least as accurate as other alternatives in a wide variety of settings, including politics [6], business [11, 42], disease surveillance [38], and entertainment [34]. The New York Times has cited prices from the popular Dublin-based market Intrade in discussions of upcoming elections,[1] and the North American Derivatives Exchange has proposed running real-money prediction markets for political events.[2] Extensive prior work has studied markets' abilities to aggregate information relevant to the value of a given security in both lab and field experiments [35–37] and at theoretical equilibria such as rational expectations equilibria [39], competitive equilibria [32, 43], and game-theoretic equilibria [10, 24, 31]. However, there has been little research studying the problem of designing securities to aggregate information relevant to externally specified events.

We develop a formal framework in which to study the design of *informative* securities for predicting the likelihood of events of interest. Loosely speaking, we consider a set of securities informative if, given their prices, it is possible to calculate the posterior probabilities of the events as if we were given the traders' private information. This suggests two requirements:

1. The price of each security offered should converge to the expected value of the security conditioned on the traders' pooled information.

---

[*]This research was partially supported by the National Science Foundation under grants IIS-1054911 and CCF-0953516. Any opinions, findings, conclusions, or recommendations expressed in this material are those of the authors alone.

[1]http://thecaucus.blogs.nytimes.com/2012/01/20/gingrich-gets-a-boost-on-intrade/

[2]http://www.freakonomics.com/2012/02/02/the-politics-of-political-prediction-markets/

2. It should be possible to uniquely map the final security prices to estimates of the likelihood of each event of interest.

Prior work [31] has described when a given security meets the first condition in perfect Bayesian equilibria. We focus on the design of securities that meet both conditions with respect to a specified set of events. This theoretical investigation offers insight on how to design prediction markets in practice to produce more accurate forecasts.

Poor security design can stop participants from aggregating their information, as the following example derived from Geanakoplos and Polemarchakis [18] shows.

**Example 1.** *Consider a market offering a single security worth $1 if a particular candidate wins the presidential election and $0 otherwise. The market has two participants: a political analyst in Washington and an Iowa caucus-goer who is well-informed on local politics. The analyst understands the importance of Iowa on the campaign and knows whether a win or loss there will mean the candidate is elected. The caucus-goer, on the other hand, knows whether the candidate will win or lose the caucus, but not its broader effect.*

*This situation can be described by defining four states of the world, $\omega_1$, $\omega_2$, $\omega_3$, and $\omega_4$:*

|  | | Iowa | |
|---|---|---|---|
|  | | Wins | Loses |
| General Election | Wins | $\omega_1$ | $\omega_2$ |
|  | Loses | $\omega_3$ | $\omega_4$ |

*The analyst knows if the true state of the world is on the diagonal or not (the effect of the caucus) and the caucus-goer knows which column the true state is in (the results of the caucus). If they could reveal their private information they would learn the true state of the world, $\omega^*$. But with a uniform prior over the state space, both think the likelihood of election is $1/2$ and value the security at $0.50 no matter what their private information is, since every signal contains a state where the candidate wins the election and another where the candidate loses. This prevents either from inferring the other's signal, and no aggregation occurs.*

The traders in Example 1 are unable to effectively express their private information and determine the correct value for the security, even though they have between them the knowledge to do so. But even when a market determines the correct value for its securities nothing may be learned. A constant-valued security worth $1 regardless of the outcome is, for example, always priced correctly, but tells us nothing about the likelihood of future events.

In practice many prediction markets offer multiple securities, like a security that pays $1 if and only if a candidate wins Iowa and another that pays $1 if and only if the candidate wins the general election. These two securities together would allow the analyst and caucus-goer to aggregate their information. The caucus-goer could first participate in the Iowa market. Then, having observed the price change in the Iowa market, the analyst would have the information he needs to participate in the general election market, revealing his own private information. Furthermore, the market organizer would be able to infer the probability of election by observing their trades. In short, these securities together are what we call *informative* on the candidate's election. We show that multiple securities are necessary for informativeness in some cases. Using too many securities, however, is bad design.

The design questions we examine are also relevant for *combinatorial prediction markets* over exponentially large outcome spaces, like the 9.2 quintillion outcomes for the NCAA tournament [44], the over $2^{50}$ ways for states to vote in the U.S. presidential election, and the $n!$ rankings for a competition with $n$ candidates. Offering a security for each state would be technically informative but practically unmanageable. Prior work has frequently considered markets with securities that correspond to natural events of interest, such as horse $A$ beating horse $B$ in a race [3, 9], but these securities may or may not be informative.

This paper considers the design of informative markets in three settings:

- In Section 5.2, we characterize securities that are *always informative*, revealing traders' information regardless of their signal structure. Complete markets, which offer one Arrow-Debreu security associated with each state of the world, are always informative for any set of events because they reveal traders' posterior distribution over the state space, but are typically too large to run in practice. We show that even if only one event is of interest, prohibitively large numbers of securities may be required to guarantee informativeness.

- Offering a large number of securities is undesirable from a practical point of view. In Section 5.3, we show that if the market designer knows the traders' signal structure, a *single* informative security can be constructed to reveal the traders' posterior distribution.

- Real markets have multiple securities, and a single security acting as a summary statistic for a complex market is unlikely to be considered natural. In Section 5.4 we consider design as an optimization problem where a designer attempts to find the fewest securities that are informative for some given events of interest but is constrained to select its securities from a predefined set. This predefined set of securities could be interpreted as the set of securities that are natural. Solving this optimization problem perfectly is NP-hard, even when the set of securities is restricted to those paying $0 or $1 in each state of the world (like the securities in the above examples).

These results suggest, unsurprisingly, that deployed prediction markets are likely not revealing *all* their participants' information and that better design may improve upon their observed efficacy. The idea of partial informativeness, which may better describe prediction markets in practice, is discussed further in the conclusion. However, only by understanding the limits and opportunities of total informativeness, as discussed in this paper, can we understand the value and interest of partial informativeness.

## 2 RELATED WORK

There is a rich literature on the informational efficiency of markets, including theoretical work on the existence and characteristics of rational expectations equilibria [4, 21, 39] and empirical studies of experimental markets [35, 36]. Here we review only the most relevant theoretical work that either focuses on the dynamic process of information aggregation or takes a security design perspective.

The early theoretical foundations of information aggregation were laid by Aumann [5], who initiated a line of research on common knowledge, establishing a solid foundation to the understanding of the phenomenon of consensus. An event $E$ is said to be *common knowledge* among a set of agents if every agent knows $E$, and every agent knows that every other agent knows $E$, ad infinitum. Aumann proved that if two rational agents have the same prior and their posterior probabilities for some event are common knowledge, then their posterior probabilities must be equal. This result was repeatedly refined and extended [18, 28, 29], and Nielsen et al. [30] showed that if $n$ agents with the same prior but possibly different information announce their beliefs about the expectation of some random variable, their conditional expectations eventually are equal.

This line of work suggests that agents will reach consensus, but says nothing about whether such a consensus fully reveals agents' information. Feigenbaum et al. [14] studied a particular model of prediction markets (a Shapley-Shubik market game [41]) in which traders' information determines the value of a security. They characterized the conditions under which the market price of the security converges to its true value under the assumption that traders are non-strategic and honestly report their expectations.

The assumption that traders are non-strategic is arguably unrealistic. Ostrovsky [31] examined information aggregation in markets with strategic, risk-neutral traders. He considered two market models, market scoring rules [22, 23] and Kyle's model [25]. For both, he showed that a separability condition is necessary for the market price of a security to always converge to the expected value of the security conditioned on all information in every perfect Bayesian equilibrium. This condition is discussed extensively in Section 4. Iyer et al. [24] extended this model to risk-averse agents and identified a smoothness condition on the price in the market that ensures full information aggregation.

The work of Feigenbaum et al. [14], Ostrovsky [31], and Iyer et al. [24] focuses on understanding the aggregation of information relevant to the value of a given, fixed security. In contrast, this paper studies how to design securities to infer the likelihood of some events of interest. There have been other papers on security design. Pennock and Wellman [33], for example, examine the conditions under which an incomplete market with a compact set of securities allows traders to hedge any risk they have (and hence is "operationally complete"). Their work considers competitive equilibria, while our work focuses on information aggregation at game-theoretic equilibria of the market.

## 3 THE MODEL

In this section, we describe our model of traders' information and the market mechanism. Our model closely follows Ostrovsky [31], but is generalized to handle a *vector* of securities (often simply referred to as a set of securities) instead of a single security.

### 3.1 Modeling Traders' Information

We consider $n$ traders, $1, \cdots, n$, and a finite set $\Omega$ of mutually exclusive and exhaustive states of the world. Traders share a common knowledge prior distribution $P_0$ over $\Omega$. Before the market opens Nature draws a state $\omega^*$ from $\Omega$ according to $P_0$ and traders learn some information about $\omega^*$ that, following Aumann [5], is based on partitions of $\Omega$. A *partition* of a set $\Omega$ is a set of nonempty subsets of $\Omega$ such that every element of $\Omega$ is contained in exactly one subset. For example, $\{\{A, B\}, \{C\}, \{D\}\}$ and $\{\{A, D\}, \{B, C\}\}$ are both partitions of $\{A, B, C, D\}$. We assume that every trader $i$ receives $\Pi_i(\omega^*)$ as their private signal, where $\Pi_i(\omega)$ denotes the element of the partition $\Pi_i$ that contains $\omega$. In other words, trader $i$ learns that the true state of the world lies in the set $\Pi_i(\omega^*)$.

We refer to the vector $\Pi = (\Pi_1, \cdots, \Pi_n)$ as the traders' *signal structure*, which is assumed to be common knowledge for all traders. The *join* of the signal structure, denoted join($\Pi$), is the coarsest common refinement of $\Pi$, that is, the partition with the smallest number of elements satisfying the property that for any $\omega_1$ and $\omega_2$ in the same element of the partition, $\Pi_i(\omega_1) = \Pi_i(\omega_2)$ for all $i$. For example, the join of the partitions $\{\{A,D\},\{B,C\}\}$ and $\{\{A,C,D\},\{B\}\}$ is $\{\{A,D\},\{B\},\{C\}\}$. The join is unique. We use $\Pi(\omega)$ to denote the element of the join containing $\omega$. Note that if two states appear in the same element of the join, no trader can distinguish between these states.

### 3.2 Market Scoring Rules

The market mechanism that we consider is a market scoring rule [22, 23]. We will describe a market scoring rule as a mechanism that allows traders to sequentially report their probability distributions or expectations. While focusing on market scoring rules may seem restrictive, market scoring rules are surprisingly general. In particular, any market scoring rule that allows traders to report probability distributions over $\Omega$ has an equivalent implementation as a cost-function-based market where the mechanism acts as an automated market maker who sets prices for $|\Omega|$ Arrow-Debreu securities, one for each state and taking value 1 in that state and 0 otherwise, and is willing to buy and sell securities at the set prices [8, 22]. This result can easily be extended to general scoring rules by applying the results of Abernethy and Frongillo [1, 2]. In particular, their results imply that any market scoring rule that allows traders to report their expectations has an equivalent implementation as a cost-function-based market that allows traders to trade securities with the market maker. Thus, without loss of generality, our model and analysis are presented for market scoring rules.

Before describing the market scoring rule mechanism, we first review the idea of a strictly proper scoring rule. *Scoring rules* are most frequently used to evaluate and incentivize probabilistic forecasts [16, 19], but can also be used to elicit the mean or other statistics of a random variable [26]. The scoring rules that we consider will be used to elicit the mean of a *vector* of random variables [40]. Let $\mathbf{X} = (x_1, \cdots, x_m)$ be a vector of bounded real-valued random variables. A scoring rule $s$ maps a forecast $\vec{y}$ in some convex region $\mathcal{K} \subseteq \mathbb{R}^m$ (e.g., the probability simplex in the case of probabilistic forecasts) and a realization of $\mathbf{X}$ to a score $s(\vec{y}, \mathbf{X}(\omega))$ in $\mathbb{R}$.[3] A scoring rule for eliciting an expectation is said to be *proper* if a risk neutral forecaster who believes that the true distribution over states $\Omega$ is $P$ maximizes his expected score by reporting $\vec{y} = \mathrm{E}_P[\mathbf{X}]$, that is, if $\mathrm{E}_P[\mathbf{X}] \in \arg\max_{\vec{y} \in \mathcal{K}} \sum_{\omega \in \Omega} P(\omega) s(\vec{y}, \mathbf{X}(\omega))$. (For random vectors $\mathbf{X}$, we use $\mathrm{E}_P[\mathbf{X}]$ to denote the expected value $\Sigma_{\omega \in \Omega} P(\omega) \mathbf{X}(\omega)$.) A scoring rule is *strictly proper* if $\mathrm{E}_P[\mathbf{X}]$ is the unique maximizer.

One common example of a strictly proper scoring rule is the Brier scoring rule [7], which is based on Euclidean distance and can be written, for any $b > 0$, as $s(\vec{y}, \mathbf{X}(\omega)) = -b \sum_{j=1}^m (y_j - x_j(\omega))^2 = -b||\vec{y} - \mathbf{X}(\omega)||^2$.

Strictly proper scoring rules incentivize myopic traders to report truthfully, but do not provide a mechanism for aggregating predictions from multiple traders. Hanson [22, 23] introduced *market scoring rules* to address this problem. A market scoring rule is a sequentially shared strictly proper scoring rule.

Formally, let $\mathbf{X}$ be a vector of random variables.[4] The market operator specifies a strictly proper scoring rule $s$ and chooses an initial prediction $\vec{y}_0$ for the expected value of $\mathbf{X}$; when there is a known common prior $P_0$, it is most natural to set $\vec{y}_0 = \mathrm{E}_{P_0}[\mathbf{X}]$. The market opens with initial prediction $\vec{y}_0$, and traders take turns submitting predictions. The order in which traders make predictions is common knowledge. Without loss of generality, we assume that traders $1, 2, \cdots, n$ take turns, in order, submitting predictions $\vec{y}_1, \vec{y}_2, \cdots, \vec{y}_n$, then the process repeats and the traders, in the same order, submit predictions $\vec{y}_{n+1}, \vec{y}_{n+2}, \cdots, \vec{y}_{2n}$. Traders repeat this process an infinite number of times before the market closes and Nature reveals $\omega^*$. Each trader then receives a score $s(\vec{y}_t, \mathbf{X}(\omega^*))$ for each prediction made at some time $t$, but must pay $s(\vec{y}_{t-1}, \mathbf{X}(\omega^*))$, the score of the previous trader. The total payment to trader $i$ (which may be negative) is then $\sum_{t=0}^\infty s(\vec{y}_{tn+i}, \mathbf{X}(\omega^*)) - s(\vec{y}_{tn+i-1}, \mathbf{X}(\omega^*))$.

### 3.3 Modeling Traders' Behavior

Together, the traders, state space, signal structure, security vector, and market scoring rule mechanism define an extensive form game with incomplete information. We consider Bayesian traders either acting in perfect Bayesian equilibrium or behaving myopically in this game. A perfect Bayesian equilibrium is a subgame perfect Bayesian Nash equilibrium. Loosely speaking, at a perfect Bayesian equilibrium, it must be the case that each player's strategy is optimal (i.e., maximizes expected utility) given the player's beliefs and the strategies of other players at any stage of the game, and that players' beliefs are derived from strategies using Bayes' rule whenever possible. See González-Díaz and Meléndez-Jiménez [20] for a more formal description.

---

[3]Technically, the region $\mathcal{K}$ should include the convex hull of the possible realizations of $\mathbf{X}$, a set equivalent to the possible expected values of $\mathbf{X}$. A full discussion of this and other properties of scoring rules is beyond the scope of this paper, but interested readers can see Savage [40].

[4]Typically market scoring rules are used for probabilistic forecasts in which case $\mathbf{X}$ would be a vector of indicator random variables, but this need not be the case.

Perfect Bayesian equilibria can be difficult to compute and it is an open question whether they always exist in prediction markets, although in some special cases they do [10]. An alternative is to consider *myopic* Bayesian traders who simply maximize their expected payoff for the current round. Since strictly proper scoring rules myopically incentivize honest reports, these traders report their current posteriors each time they make a prediction.

## 4 AGGREGATION

*Separability* is used to characterize the conditions under which securities aggregate information about their own values. Building on ideas from DeMarzo and Skiadas [12, 13], Ostrovsky [31] characterized separability for a single security. He showed that in every perfect Bayesian equilibrium market prices will, in the limit, reflect the value of the security as if traders had revealed their private signals if and only if the security is separable. If a security is not separable, then there always exist priors and equilibrium strategies where no information aggregation occurs.

In this section, we generalize these prior definitions to multiple securities and arbitrary signal structures. Ostrovsky assumed a restricted class of signal structures without loss of generality, and these generalizations are uninteresting when only considering aggregation. They will be necessary to discuss informativeness, however, as the results of the next section demonstrate. We then restate Ostrovsky's equilibrium aggregation result in this setting. As previously discussed, perfect Bayesian equilibrium may or may not exist in prediction markets, and we also adapt and formalize prior work on information aggregation to show separability is also the necessary and sufficient condition for myopic traders to always aggregate their information.

Informative markets require separable securities. If a market uses separable securities then both traders in perfect Bayesian equilibrium and Bayesian traders acting myopically will, in the limit, value the security as if their private signals were revealed, and this allows a market designer to directly infer the likelihood of his events of interest from the securities' value. If a set of non-separable securities were used then the market designer could be required instead to perform additional inference and know the prior and traders' strategies.

As mentioned, we say a market aggregates information if, in the limit as time goes to infinity, the value of the securities approaches their value conditional on all the traders' private signals. Since each trader $i$ receives the signal $\Pi_i(\omega^*)$, their pooled signal is $\bigcap_i \Pi_i(\omega^*) = \Pi(\omega^*)$.

**Definition 1** (Aggregation). *Information is aggregated with respect to a set of securities $\mathbf{X}$, signal structure $\Pi$, and common prior $P_0$, if the sequence of predictions $\vec{y}_0, \vec{y}_1, \vec{y}_2, \cdots$ converges in probability to the random vector $E_{P_0}[\mathbf{X}|\Pi(\omega^*)]$.*

A set of securities is separable if and only if the traders only agree on their value when it reflects their pooled information. That is, for any prior distribution there must be at least one trader whose private information causes them to dissent from a consensus, unless that consensus is the traders' collective best estimate.

**Definition 2** (Separability). *A set of securities $\mathbf{X}$ is* non-separable *under partition structure $\Pi$ if there exists a distribution $P$ over $\Omega$ and vector $\vec{v}$ such that $P(\omega) > 0$ on at least one state $\omega \in \Omega$ in which $E_P[\mathbf{X}|\Pi(\omega)] \neq \vec{v}$, and for every trader $i$ and state $\omega, P(\omega) > 0$,*

$$E_P[\mathbf{X}|\Pi_i(\omega)] = \frac{\sum_{\omega' \in \Pi_i(\omega)} P(\omega')\mathbf{X}(\omega')}{\sum_{\omega' \in \Pi_i(\omega)} P(\omega')} = \vec{v}. \quad (1)$$

*If a security is not non-separable then it is* separable.

Here the vector $\vec{v}$ represents a possible consensus, only agreed upon if there is no alternative when the securities are separable. Separability is a property of the entire set of securities, as Example 2 demonstrates.

**Example 2.** *Let $\Omega = \{\omega_1', \omega_2^*, \omega_3, \omega_4', \omega_5^*, \omega_6\}$. Two traders have partitions as follows:*

$$\Pi_1 = \{\{\omega_1', \omega_2^*, \omega_3\}, \{\omega_4', \omega_5^*, \omega_6\}\}$$
$$\Pi_2 = \{\{\omega_1', \omega_5^*\}, \{\omega_3, \omega_4'\}, \{\omega_2^*, \omega_6\}\}$$

*and there are two securities: $x^*$ with value one when $\omega_2^*$ or $\omega_5^*$ occurs and zero otherwise, and $x'$ with value one when $\omega_1'$ or $\omega_4'$ occurs and zero otherwise.*

*Both securities are individually non-separable with respect to $\Pi$. If the prior $P$ is uniform over $\omega_1', \omega_2^*, \omega_5^*,$ and $\omega_6$, then $E_P[x^*|\Pi_i(\omega)] = 1/2$ for $i \in \{1, 2\}$ and all $\omega$ such that $P(\omega) > 0$. Similarly, if $P$ is uniform over $\omega_1', \omega_3, \omega_4',$ and $\omega_5^*$, then $E_P[x'|\Pi_i(\omega)] = 1/2$ for $i \in \{1, 2\}$ and all $\omega$ such that $P(\omega) > 0$. The join of traders' partitions, however, consists of singletons. Hence, both $E_P[x^*|\Pi(\omega)]$ and $E_P[x'|\Pi(\omega)]$ have value 0 or 1, not 1/2, for all $\omega$.*

*But taken together the set of securities is separable with respect to $\Pi$. Given any prior distribution $P$*

and a state $\omega$, trader 2 either identifies $\omega$ with certain, which happens when $P$ assigns 0 probability to the other state in its signal $\Pi_2(\omega)$, or assigns positive probability to both states in $\Pi_2(\omega)$. In the former case, $E_P[\mathbf{X}|\Pi_2(\omega)] = E_P[\mathbf{X}|\Pi(\omega)]$. In the latter case, trader 2's expected value for the securities is positive for both when $\omega \in (\omega_1', \omega_5^*)$, positive for only $x'$ when $\omega \in (\omega_3, \omega_4')$, and positive for only $x^*$ when $\omega \in (\omega_2', \omega_6)$. If the set of securities is non-separable there must exist a distribution $\tilde{P}$ and a vector $\vec{v}$ such that $\vec{v} \neq E_{\tilde{P}}[\mathbf{X}|\Pi(\tilde{\omega})]$ for some state $\tilde{\omega} \in \{\omega|P(\omega) > 0\}$ and $E_{\tilde{P}}[X|\Pi_2(\omega)] = \vec{v}$ for any state $\omega \in \{\omega|P(\omega) > 0\}$. This is possible only when $\tilde{P}$ assigns positive probability to the two states in $\Pi_2(\tilde{\omega})$ and 0 probability for all other states because each signal of player 2 has a distinct expectation of the securities. Given such a $\tilde{P}$, however, trader 1 always uniquely identifies the true state and has the correct expectation of the securities. Hence, the set of securities is separable with respect to $\Pi$.

### 4.1 Aggregation

Separability is a necessary and sufficient property for aggregation in two natural cases.

**Theorem 1** (Equilibrium Aggregation, Ostrovsky [31])**.** *Consider a market with securities $\mathbf{X}$ and traders with signal structure $\Pi$. Information is aggregated in every perfect Bayesian equilibrium of this market if and only if the securities $\mathbf{X}$ are separable under $\Pi$.*

**Theorem 2** (Myopic Aggregation)**.** *Consider a market with securities $\mathbf{X}$ and myopic traders with signal structure $\Pi$. Information is aggregated in finite rounds if and only if the securities $\mathbf{X}$ are separable under $\Pi$.*

Ostrovsky [31] proved a special case of Theorem 1 for markets with one security. Theorem 1 stated above accommodates any finite set of securities and is proved using a simple extension of Ostrovsky's proof. Specifically, the proof shows that traders' sequences of predictions at any perfect Bayesian equilibrium are bounded martingales and must converge. Separability implies that if information is not aggregated in the limit, there exists an agent who can make an arbitrarily large profit by deviating from his equilibrium strategy, a contradiction to traders being in equilibrium.

The proof of Theorem 2 makes use of prior work on convergence to common knowledge (particularly Geanakoplos [17]) and shows not only that myopic traders' sequences of predictions are bounded martingales but also that they must converge to the same random vector in a finite number of periods. Then, by separability, it is shown that this consensus prediction must equal $E[X|\Pi(\omega^*)]$, implying aggregation. A full proof appears in the appendix of the long version of this paper, available on the authors' websites.

If the securities are not separable then there exists a distribution $P$ satisfying (1) in the definition of separability. Letting this distribution be the prior, a perfect Bayesian equilibrium is simply for traders to report the common consensus value, not allowing any meaningful Bayesian updating and preventing aggregation from occurring. Myopic traders are constrained to report this same value.

## 5 SECURITY DESIGN

In this section we discuss the design of informative markets. While separability is a sufficient and necessary condition for aggregation in two natural settings, it only implies the value of the securities reflects all the traders' private information, not that the market designer can use this value to infer that private information or the likelihood of the events of interests. We define informative securities as securities that are both separable and allow for the likelihood of the events of interest to be inferred directly from their value.

As we will show, complete markets are always informative, but deployed prediction markets are rarely complete. These markets require too many securities to be practical, and their securities present challenges for traders. A prediction market for the U.S. presidential election, for example, may need one state per outcome in the electoral college. This is over $2^{50}$ states and requires traders to bid on securities like "The President wins Ohio, not Florida, Illinois, not Indiana ..." Even if alternative bidding methods were developed, traders would still be required to review the value of each security for aggregation to be formally implied. This is impractical, and so we consider good designs as those using a few natural securities. We first discuss the design requirements of markets that are always informative, and markets that are informative for a particular signal structure. The latter market allows a single security to be informative on any set of events, but arguably appears "unnatural." To describe the challenges of designing using only natural securities we then consider a constrained design process instead, where the market designer is restricted to an arbitrary subset of (possibly natural) securities.

### 5.1 Informative Securities

Informally, we would like to say that a market's securities are *informative* on a set of events with respect to a signal structure if the market organizer learns the likelihood of the events as if it knew all the traders' private signals. Assuming the values of the securities reflect traders' pooled information, if the likelihood of the events is unambiguously implied from these values then functionally all the private signals are revealed. We call this latter property *distinguishability*.

**Definition 3** (Distinguishability). *Let $\Pi$ be a signal structure over states $\Omega$ and $P_{\text{join}(\Pi)}$ be the set of all probability distributions over $\Omega$ that assign positive probability only to a subset of states in one element of $\text{join}(\Pi)$ (i.e., a trader's possible posteriors after aggregation). A set of securities $\mathbf{X}$ on $\Omega$ distinguishes a set of events $\mathcal{E}$ with respect to $\Pi$ if and only if for any $P, P' \in P_{\text{join}(\Pi)}, E_P[\mathbf{X}] = E_{P'}[\mathbf{X}]$ implies $P(E) = P'(E), \forall E \in \mathcal{E}$.*

Equivalently a set of securities distinguishes a set of events if there exists a function from the securities' values to the likelihood of the events. When a set of securities is both separable and distinguishable we describe it as *informative*.

**Definition 4** (Informativeness). *A set of securities $\mathbf{X}$ is informative on a set of events $\mathcal{E}$ with respect to a signal structure $\Pi$ if and only if $\mathbf{X}$ both distinguishes $\mathcal{E}$ and is separable with respect to $\Pi$.*

Informativeness is a strong condition. Even if securities are not informative it might be possible for a market designer to infer some information from the market, or for the market to be described as partially informative. Generalizing our framework to account for partial aggregation would be an interesting line of future work.

## 5.2 Always Informative Securities

We first address the problem of designing a set of securities that is informative on a set of events with respect to *any* signal structure. We call such securities *always informative*. These securities may be of practical interest if the market designer is unsure of the traders' signal structure; using a set of always informative securities implies aggregation will occur no matter what the true signal structure is.

A market is said to be *complete* if by trading securities, agents can freely transfer wealth across states [27]. Rigorously, consider the set of securities that contains a constant payoff security plus all of the securities offered by a market. The market is complete if and only if this set includes $|\Omega|$ linearly independent securities. The most common is a market with $|\Omega|$ Arrow-Debreu securities, each associated with a different state of the world, taking value 1 on that state and 0 everywhere else. For an overview of complete markets, see Flood [15] or Mas-Colell et al. [27].

Complete markets are theoretically appealing because they allow traders to express any information about their beliefs. We formalize this well-known idea in our framework in the following proposition.

**Proposition 1.** *A market over state space $\Omega$ with securities $\mathbf{X}$ is complete if and only if for all distinct probability distributions $P$ and $P'$ over $\Omega$, $E_P[\mathbf{X}] \neq E_{P'}[\mathbf{X}]$.*

*Proof.* Let $M$ be a matrix containing the payoffs of $\mathbf{X}$, with one row for each outcome and one column for each security. The element at row $i$ and column $j$ of M takes value $\mathbf{x}_j(\omega_i)$. Consider a probability distribution $P$ represented as a row vector so, $PM = \mathrm{E}_P[\mathbf{X}]$. The system of linear equations

$$P'M = \mathrm{E}_P[\mathbf{X}] \qquad \sum_{\omega \in \Omega} P'(\omega) = 1$$

has a unique solution $P' = P$ if and only if the matrix $M'$, which is $M$ augmented by a column of 1s to represent the summation constraint, has rank $|\Omega|$.

If the market is complete, $M'$ has this rank so any distinct probability distribution has distinct expectation.

Now assume $\mathrm{E}_P[\mathbf{X}] \neq \mathrm{E}_{P'}[\mathbf{X}], \forall P \neq P'$, and, for a contradiction, that the market is not complete. Then the system of equations has at least two solutions, one of which is the probability distribution $P$ and a distinct solution $Q$, such that $PM = QM = \mathrm{E}_P[X]$. Let $U$ be the uniform distribution over $\Omega$. Then there exists $c > 0$ such that $(1-c)U + cQ$ is a probability distribution (since $Q$ satisfies $\sum_{\omega \in \Omega} Q(\omega) = 1$). Moreover, $(1-c)U + cP$ is also a probability distribution and

$$\big((1-c)U + cP\big)M = \big((1-c)U + cQ\big)M,$$

contradicting $\mathrm{E}_P[\mathbf{X}] \neq \mathrm{E}_{P'}[\mathbf{X}], \forall P \neq P'$. Thus, the market must be complete. $\square$

This expressiveness is a necessary and sufficient condition for the likelihood of every event to be inferred, and suggests an alternative characterization of complete markets as those markets that are always informative on every event.

**Theorem 3.** *A market is always informative on every event $E$ with respect to every signal structure $\Pi$ if and only if it is complete.*

*Proof.* Distinguishing every event $E$ is equivalent to distinguishing each state of the world $\omega \in \Omega$; the latter are also events and so must be distinguished, and if each is distinguished then the likelihood of any event can be inferred. Proposition 1 shows that completeness is a necessary and sufficient condition for distinguishing each state of the world.

It remains to show that complete markets are also separable with respect to any signal structure $\Pi$. Assume,

for a contradiction, there exists a signal structure $\Pi$ and a complete market with securities $\mathbf{X}$ such that $\mathbf{X}$ is non-separable with respect to $\Pi$. Since $\mathbf{X}$ is non-separable there must exist distinct probability distributions $P$ and $P'$ over $\Omega$ such that $\mathrm{E}_P[\mathbf{X}] = \mathrm{E}_{P'}[\mathbf{X}]$; but by Proposition 1, in a complete market this equality only holds if $P = P'$, a contradiction. So complete markets are separable with respect to any signal structure and always distinguish every event, implying they are always informative on every event. $\square$

Complete markets are often impractical, but rarely is every event of interest. Even if a single event is of interest, however, as many securities as almost half the states in the market may be required to create an always informative market. We let $\bar{E}$ denote the complement of $E$.

**Theorem 4.** *Any market that is always informative on an event $E$ must have at least $\min(|E|, |\bar{E}|) - 1$ linearly independent securities.*

*Proof.* Let $\mathbf{X}$ be a set of securities, fewer than $\min(|E|, |\bar{E}|)) - 1$ of which are linearly independent, and assume, for a contradiction, that $\mathbf{X}$ is always informative on $E$. Restricting attention to states in $E$, the argument from Proposition 1 implies this market has too few securities to distinguish every probability distribution over $E$ and there exist probability distributions $P_E$ and $P'_E$ such that $E_{P_E}[\mathbf{X}] = E_{P'_E}[\mathbf{X}]$. Let the difference between these distributions be the vector $\Delta_E = P_E - P'_E$, and define vectors $\Delta_E^+$ and $\Delta_E^-$ such that $\Delta_E^+(\omega) = \max(0, \Delta_E(\omega))$ and $\Delta_E^-(\omega) = \min(0, \Delta_E(\omega))$. Since $\Delta_E$ is the difference of two probability distributions with the same expected value,

$$\sum_{\omega \in E} \frac{\Delta_E^+(\omega)}{||\Delta_E^+||_1} \mathbf{X}(\omega) = \sum_{\omega \in E} \frac{-\Delta_E^-(\omega)}{||\Delta_E^-||_1} \mathbf{X}(\omega). \quad (2)$$

That is, $\frac{\Delta_E^+(\omega)}{||\Delta_E^+||_1}$ and $\frac{-\Delta_E^-(\omega)}{||\Delta_E^-||_1}$ are disjoint probability distributions over states in $E$ with the same expected value, and the same argument can be made, *mutatis mutandi* for two such probability distributions over states in $\bar{E}$. Let these distributions over $E$ be $Q_E$ and $Q'_E$, and the ones over $\bar{E}$ be $Q_{\bar{E}}$ and $Q'_{\bar{E}}$. Although we have been referring to these as distributions over $E$ and $\bar{E}$ we consider them to be distributions over $\Omega$ that assign zero probability to all states not previously included in the distributions, and we will use these names to stand for both these distributions and the states they assign positive probability to to reduce notation.

Now suppose there are two traders with signal structure

$$\Pi_1 = \{\{Q_E, Q_{\bar{E}}\}, \{Q'_E, Q'_{\bar{E}}\}\}$$
$$\Pi_2 = \{\{Q_E, Q'_{\bar{E}}\}, \{Q'_E, Q_{\bar{E}}\}\}$$

and prior

$$P_0 = \frac{Q_E + Q'_E + Q_{\bar{E}} + Q'_{\bar{E}}}{4}.$$

Each trader's expectation conditional on any signal is the same since $\mathrm{E}_{Q_E}[\mathbf{X}] = \mathrm{E}_{Q'_E}[\mathbf{X}]$ and $\mathrm{E}_{Q_{\bar{E}}}[\mathbf{X}] = \mathrm{E}_{Q'_{\bar{E}}}[\mathbf{X}]$ and each signal contains one distribution over states in $E$ and another over states in $\bar{E}$. But the join of the signal structure is $\{\{Q_E\}, \{Q'_E\}, \{Q_{\bar{E}}\}, \{Q'_{\bar{E}}\}\}$, and if $\mathbf{X}$ is separable with respect to $\Pi$ the expectation conditional on any such element must also, then, be the same. This implies $E_{Q_E}[\mathbf{X}] = E_{Q_{\bar{E}}}[\mathbf{X}]$, but by construction $Q_E(E) \neq Q_{\bar{E}}(E)$, so if $\mathbf{X}$ is separable with respect to $\Pi$ it does not distinguish $E$, contradicting our assumption that $\mathbf{X}$ is always informative on $E$. $\square$

This result demonstrates the need for a market designer to allow traders to express information it finds uninteresting. It also suggests that, in practice, few markets are acquiring all of their participants' information. This is unsurprising, but we think better designs will extract more information, and that this result shows knowledge of or assumptions about the traders' signal structure may be necessary to inform those designs.

### 5.3 Fixed Signal Structures

If the join of the traders' signal structure is known and has singleton sets for its elements, then there exists a single security that is informative on every event.

**Theorem 5.** *For any signal structure $\Pi$ such that $\mathrm{join}(\Pi)$ consists only of singleton sets there exists a security $\mathbf{x}$ that is informative on every event $E$ with respect to $\Pi$.*

The proof uses a result from Ostrovsky [31].

**Theorem 6** (Ostrovsky [31]). *Let $\Pi$ be a signal structure such that $\mathrm{join}(\Pi)$ consists of singleton sets of states, and let $x$ be a security that can be expressed as $x(\omega) = \Sigma_i f(\Pi_i(\omega))$ for an arbitrary function $f$ mapping signals to reals. Then $x$ is separable under $\Pi$.*

*Proof of Theorem 5.* To construct the security, first assign a unique identifier $s_0, s_1, s_2, \ldots$ to every signal of every trader, and define $f(s_j) = 10^j$ for all $j$. Let $S_\omega$ denote the set of indices of the identifiers corresponding to the signals of each trader for state $\omega$, i.e, corresponding to $\Pi_i(\omega)$ for each trader $i$. The security

$x(\omega) = \Sigma_{j \in S_\omega} f(s_j)$ is separable by Theorem 6. Additionally, the sum $\Sigma_{j \in J} f(s_j)$ for any $J \subset \{0, 1, 2, \ldots\}$ is unique, and each state $\omega$ has a unique associated set of signals since we assumed the join consists of singletons. This implies the value of the security for each element of the join is unique, so the security also distinguishes every event. □

The assumption that the join of traders' signal structure consists only of singleton sets is not without loss of generality. If the signal structure is known, however, the market designer can treat elements of the join as states of the world, identify the correct element of the join by running the market with a single security, then apply the prior to that element to learn the likelihood of each state as if he knew all the traders' private signals. If the prior is unknown this distribution can also be solicited from any single trader using a scoring rule.

### 5.4 Constrained Design

A single security acting as a summary statistic for an entire market is unlikely to be considered natural by any criterion. Real markets, like those on Intrade, use multiple securities. Instead of imposing our own definition of natural, in this section we consider adding a design constraint that the market's securities must be picked from a predefined set. The market designer is then challenged to find the fewest securities from this set that are informative on the events of interest with respect to the given signal structure. We call this the INFORMATIVE SET optimization problem. If the set of predefined securities is empty or has no informative subset then the problem is simply infeasible, so we assume there exists at least one such subset.

Demonstrating INFORMATIVE SET is hard would not be very interesting if exotic and unnatural securities were required for the proof. One commonly used class of securities are *event securities* which pay $1 if an event occurs and $0 otherwise. The corresponding optimization problem is INFORMATIVE EVENT SET, a restriction of INFORMATIVE SET, and even solving this restricted version of the problem is NP-hard.

**Theorem 7.** INFORMATIVE EVENT SET *is* NP-*hard*.

This immediately implies that the more general INFORMATIVE SET problem is also hard.

**Corollary 1.** INFORMATIVE SET *is* NP-*hard*.

The proof appears in the appendix and demonstrates a one-to-one correspondence between set cover instances and a minimal informative set of securities for a single fully informed trader.

The complexity of these problems suggests that while knowledge of the traders' signal structure allows for better designs, a perfect design will be intractable to compute or require additional assumptions about the relationship between traders' signal structure and the set of possible securities. Practically we can only ever hope to offer better (but not perfect) designs that extract more information from traders than current markets do. These results confirm we will always have to settle for some degree of error in our designs even if the traders' signal structure could be perfectly observed.

## 6 CONCLUSION

We developed a formal framework for the design of informative prediction markets. These markets reveal the posterior probabilities of a set of events of interest as accurately as if the traders had directly revealed their information, a commonly cited goal of prediction markets. These markets require that traders have an incentive to be accurate, that they can aggregate their information, and that the market designer can use this information to infer the likelihood of the events of interest.

Ideally informative markets would use a few natural securities. Complete markets, usually too large to be used in practice, are, however, the only markets to always be informative, regardless of the traders' signal structure. When the signal structure is known, a single security can be informative, but this security may appear strange and unintuitive to traders. Finding the smallest informative set of natural securities is computationally hard in general.

Real-world prediction markets do typically offer small numbers of simple and natural securities, and have been shown to aggregate information effectively in practice. This is not in contrast to our results, which only consider whether a market reveals *all* of the traders' private information. Our results demonstrate, however, the importance of security design and suggest that better designs that extract *more* of the traders' information are possible.

We hope this paper will allow future research on *partial* aggregation, like that which occurs in practice. Future work might also consider the effect of alternative communication channels outside the market's securities, like comments on Intrade, that allow traders to explain the reasoning behind their predictions. The development of this line of research is crucial to understanding why prediction markets work and how to make them work better.